\documentclass[conference]{IEEEtran}
\IEEEoverridecommandlockouts
\usepackage{cite}
\usepackage{amsmath,amssymb,amsfonts}
\usepackage{algorithmic}
\usepackage{graphicx}
\usepackage{mathtools}
\usepackage{textcomp}
\usepackage{url}
\usepackage{xcolor}
\def\BibTeX{{\rm B\kern-.05em{\sc i\kern-.025em b}\kern-.08em
    T\kern-.1667em\lower.7ex\hbox{E}\kern-.125emX}}
\begin{document}

\title{Throughput and Capacity Evaluation of 5G New Radio Non-Terrestrial Networks with LEO Satellites}

\author{\IEEEauthorblockN{Jonas Sedin}
\IEEEauthorblockA{\textit{Ericsson Research} \\
jonas.sedin@ericsson.com}
\and
\IEEEauthorblockN{Luca Feltrin}
\IEEEauthorblockA{\textit{Ericsson Research} \\
luca.feltrin@ericsson.com}
\and
\IEEEauthorblockN{Xingqin Lin}
\IEEEauthorblockA{\textit{Ericsson Research} \\
xingqin.lin@ericsson.com}
\thanks{
This is the authors' version of an article that has been published accepted by IEEE Globecom 2020. \textcopyright 2000 IEEE. Personal use of this material is permitted. Permission from IEEE must be obtained for all other uses, in any current or future media, including reprinting/republishing this material for advertising or promotional purposes, creating new collective works, for resale or redistribution to servers or lists, or reuse of any copyrighted component of this work in other works.}
}

\maketitle

\begin{abstract}
A non-terrestrial network (NTN), a term coined by the 3rd Generation Partnership Project (3GPP), refers to a network utilizing airborne or spaceborne payload for communication. The use of NTN has the potential of facilitating providing connectivity to underserved areas. This has motivated the work in 3GPP on evolving the fifth generation (5G) wireless access technology, known as new radio (NR), to support NTN. The broadband opportunities promised by NTN with low Earth orbit (LEO) satellites have attracted much attention, but the performance of LEO NTN using 5G NR has not been well studied. In this paper, we address this gap by analyzing and evaluating the throughput and capacity performance of LEO NTN. The evaluation results show that the downlink capacity of a LEO satellite in S band with 30 MHz bandwidth serving handheld terminal is about 600 Mbps and the downlink capacity of a LEO satellite in Ka band with 400 MHz bandwidth serving very small aperture terminal (VSAT) is about 7 Gbps. For a LEO NTN similar to the Kuiper project proposed by Amazon, we find that, due to the large cell sizes in the LEO NTN, the area capacity density is moderate:  1 – 10 kbps/km$^2$ in the S band downlink and 14 – 120 kbps/km$^2$ in the Ka band downlink depending on latitude.
\end{abstract}


\section{Introduction}\label{Intro}
The Release 15 of the 3rd Generation Partnership Project (3GPP) represents the first release of the fifth generation (5G) mobile communications system and includes a new radio access technology known as New Radio (NR). 5G NR features spectrum flexibility, ultra-lean design, forward compatibility, low latency support, and advanced antenna technologies \cite{b1}. 5G NR evolution has been carried out to address various requirements from a wide range of industries and verticals including the support for non-terrestrial access ecosystem. Non-terrestrial networks (NTN) refer to networks utilizing airborne or spaceborne payload for communication. Examples of spaceborne platforms include low Earth orbiting (LEO) satellites, medium Earth orbiting (MEO) satellites, and geosynchronous Earth orbiting (GEO) satellites.

Interoperability has traditionally been difficult in the non-terrestrial satellite access ecosystem \cite{b2}. To overcome this, the satellite industry has joined forces with the mobile industry in 3GPP to evolve 5G standards to support satellite communications. This will allow satellite industry to benefit from the strong mobile ecosystem. A series of study items have been completed in 3GPP for supporting NTN. In Release 15, 3GPP studied the scenarios, requirements, and channel models for 5G NR based NTN \cite{b3}. In Release 16, 3GPP continued the effort by examining the solutions for adapting 5G NR to support NTN \cite{b4}. These two study items have led to an approved work item \cite{b5} that will develop standardization enhancements to evolve 5G NR to support NTN in 3GPP Release 17. Overview of the 3GPP work can be found in \cite{b6, b7}.

During the last couple of years a set of prominent LEO networks including Starlink, Kuiper and OneWeb have been proposed. These proposals feature using large LEO constellations with thousands of LEO satellites to provide global broadband access. Thus for NTN it is believed that LEO NTNs are expected to hold the greatest broadband opportunities.  However, the performance of LEO NTN has not been well understood. The objective of this paper is to address this gap by analyzing and evaluating the throughput and capacity performance of LEO NTN.

Different system-level evaluations have been performed for satellite networks in the past. In \cite{b8}, analytical expressions along with geometrical simulations were used for typical satellite orbits (LEO, MEO and GEO) to derive statistical performance quantities such as coverage areas, frequency of handovers, and link absence periods. This work proposed several techniques such as spot turnoff and frequency division to reduce interference in non-GEO systems. In \cite{b9}, the capacity of a MEO network using DVB-S2X (digital video broadcasting for second generation satellite extensions) was evaluated using a semi-dynamic simulation procedure. It was shown in \cite{b9} that close to 10 Gbps can be achieved with the use of large antenna sizes and 1.3 GHz bandwidth. In \cite{b10}, the authors focused on the assessment of optical LEO downlinks and pointed out that large networks are required to meet the requirement of latency-critical applications. The work \cite{b11} developed a statistical method to estimate the total system throughput and used it to compare the three large LEO satellite constellations: Starlink, OneWeb, and Telesat. In \cite{b12}, the author provided a preliminary evaluation of the latency performance of Starlink and concluded that a network built with a mega-LEO constellation can provide low latency performance in wide-area communication. The work \cite{b13} presented a capacity study of a LEO NTN using multi-input and multi-output (MIMO). In \cite{b14}, the authors analysed the impact of beam coverage characteristics on the performance of LEO systems.

In this paper, we study and evaluate the throughput and capacity of LEO NTN that utilizes 5G NR. We carry out system throughput simulations based on the latest developments from 3GPP including channel models, antenna models, and network scenarios. We also derive analytical expressions for LEO NTN system capacity estimation and apply them to estimate the system capacity of a mega-LEO constellation. The presented analysis and results in this paper help shed light on the use cases that can be supported by LEO NTN. 

The remainder of this paper is organized as follows. In Section \ref{SimMethodology}, we present system simulation methodology for the considered LEO NTN systems. Then in Section \ref{ConstMethodology}, we describe the evaluation methodology for estimating the system capacity density of a mega-LEO constellation. We present system simulation results in Section \ref{SimResult} and constellation capacity evaluation results in Section \ref{ConstResult}. We conclude this paper and point out future work in Section \ref{Conclusion}.

\section{System Simulation Methodology}\label{SimMethodology}
Dynamic system evaluation results presented in this paper are produced mainly based on the latest developments from 3GPP NTN studies. In this section, we describe the main evaluation assumptions and refer interested readers to the reports of 3GPP NTN study items for more details \cite{b3, b4}.

\subsection{Simulation Setup Overview}\label{SimOverview}
We consider NTN systems with frequency division duplex (FDD), which is the dominant duplexing option for satellite spectrum allocation. We consider two frequency bands for the NTN systems. For the S band, the nominal downlink and uplink carrier frequencies are both at 2 GHz, the system bandwidth is 30 MHz, and the subcarrier spacing is 15 kHz. For the Ka band, the nominal downlink and uplink carrier frequencies are at 20 GHz and 30 GHz, respectively, the system bandwidth is 400 MHz, and the subcarrier spacing is 60 kHz.

The NTN systems utilize LEO satellites with 600 km orbital height. Each LEO satellite generates multiple beams to serve terminals on the ground. The coverage area of each beam can be regarded as the coverage area of a cell as in terrestrial cellular network. The footprint of a beam is also often referred to as a spotbeam.

The simulation is performed using a dynamic simulator implemented with full 5G NR protocol stack which includes full control signaling for scheduling and hybrid automatic repeat request (HARQ) feedback, a scheduler that schedules based on received Buffer Status Reports in a proportionally fair manner, and control plane procedures such as handover through UE-based measurements reports.

\subsection{Satellites and Terminal Parameters}\label{SatParameters}

Satellite antenna is assumed to be a typical reflector antenna with a circular aperture. The corresponding normalized antenna gain pattern, denoted by $g(\theta)$, is given by
\begin{equation}
g(\theta) = \left\{\begin{matrix}
4 \left | \dfrac{J_1 \left ( \dfrac{2\pi}{\lambda}a \sin \theta \right )}{\dfrac{2\pi}{\lambda}a \sin \theta} \right |^2 & 0 < \theta \leq 90 \\ 
1 & \theta = 0 
\end{matrix}\right.
\label{eq:BeamPattern}
\end{equation}
where $\theta$ is the angle measured from the antenna boresight, $J_1(x)$ is the Bessel function of the first kind and first order, $a$ is the radius of the antenna's circular aperture, $\lambda$ is the carrier wavelength. 

An important parameter, which will be used to determine the satellite beam positions and sizes, is the half power beam width (HPBW) denoted as $\phi_{HPBW}$. It is defined as the angular width of the radiation pattern between points 3 dB down from the beam peak. The HPBW of the satellite antenna depends on the carrier wavelength and the radius of the antenna’s circular aperture, as given by \eqref{eq:BeamPattern}. For the satellite operating at S band, $\phi_{HPBW}$ = 4.41°, the effective isotropic radiated power (EIRP) is 34 dBW/MHz, and the antenna gain-to-noise-temperature (G/T) is 1.1 dB/K. For the satellite operating at Ka band, $\phi_{HPBW}$ = 1.76° the EIRP is 4 dBW/MHz, and the antenna G/T is 13 dB/K.

The S band NTN system is assumed to serve handheld user equipment (UE). The handheld UE has two cross-polarized isotropic antennas. The antenna gain of each antenna is 0 dBi. The handheld UE transmit power is 23 dBm and the antenna G/T equals -31.6 dB/K. The Ka band NTN system is assumed to serve very small aperture terminal (VSAT), which may be used, for example, in fixed wireless or marine vessels. The VSAT UE is equipped with circular polarized dish antenna. The antenna gain pattern is also given by \eqref{eq:BeamPattern} with antenna radius of 30 cm. The transmit and receive antenna gains are 43.2 dBi and 39.7 dBi, respectively. The VSAT UE transmit power is 33 dBm and the antenna G/T equals 15.9 dB/K.

\subsection{Satellite Beam Layout}\label{BeamLayout}
The satellite beams point towards the ground in a hexagonal manner similar to the canonical evaluation setup with hexagonal tessellation used for terrestrial cellular networks. The hexagonal grid is defined through the UV-plane coordinate system. The transformation from spherical coordinate system to UV-plane coordinate system is given by
\begin{equation}
\begin{matrix}
u = \sin(\theta)\cos(\omega) \\ 
v = \sin(\theta)\sin(\omega)
\end{matrix}
\label{eq:UV}
\end{equation}
where $\theta$ and $\omega$ denote azimuth angle and elevation angle, respectively. The U-axis is defined as the perpendicular line to the satellite-earth line on the orbital plane, and the straight line orthogonal to the UV plane points towards the Earth center.

To model inter-beam interference, 19 spotbeams are created, and the mapping of the spotbeams to cells is one-to-one, thus we will refer to cells instead of spotbeams. The adjacent beam spacing is computed based on the half power beam width of the satellite antenna pattern. By using the UV-plane coordinate system, a homogeneous separation between different beams can be achieved. We further assume universal frequency reuse across the 19 beams.

\begin{figure}[tbp]
\centerline{\includegraphics[width=0.7\columnwidth]{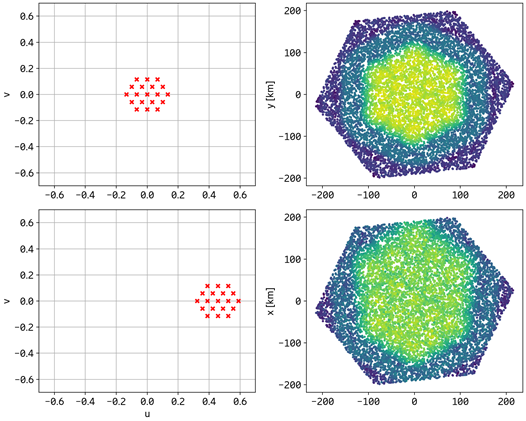}}
\caption{UV-plane pattern and normalized satellite antenna gain on the ground for a satellite at 600 km height: 90° center elevation angle in the upper figures and 60° center elevation angle in the bottom figures.}
\label{f:beams}
\vspace{-4mm}
\end{figure}

Figure \ref{f:beams} provides an illustration of the UV-plane pattern and normalized satellite antenna gain distribution on the ground. In the upper figures, the central beam center is at nadir point, i.e., 90° elevation angle. In the bottom figures, the central beam center has 60° elevation angle.

\section{Constellation Capacity Evaluation Methodology}\label{ConstMethodology}

The evaluation methodology presented in Section \ref{SimMethodology} allows for simulating part of the NTN systems. It is however not enough for estimating the capacity of a complete NTN system with a mega-LEO constellation. In order to estimate the full capabilities of a LEO satellite constellation, it is important to consider the LEO network as a whole. In this section, we present an analytical framework for estimating the system capacity of a LEO constellation.

We consider a satellite constellation formed by satellites located in a number of orbital planes, denoted by $N_p$, distributed around Earth. We consider circular orbital planes at altitude $h$. In each orbital plane, there are $N_{spp}$ satellites that are uniformly distributed. The total number of satellites in the constellation, denoted by $N_{sat}$, is then given by $N_{sat} = N_p N_{spp}$. We further assume that all the orbital planes are inclined with respect to the equatorial plane by an inclination angle denoted by $\alpha$.

The elevation angle of UE, which refers to the angle between the line pointing towards the satellite and the local horizontal plane, is denoted by $\epsilon$. Typically, a minimum elevation angle, denoted by $\epsilon_{min}$, is defined in such a way that the a UE-satellite connection is not established if the elevation angle of the UE with respect to the satellite is below $\epsilon_{min}$ due to insufficient link budget. By defining this minimum elevation angle, it is possible to define a portion of Earth surface below the satellite where a UE can connect to the satellite. We call this satellite service area as satellite footprint. We further define footprint angle, denoted by $\beta_{max}$, which is the angle between the segment from Earth center to the satellite footprint center and the segment from Earth center to any point on the satellite footprint edge. Denoting by $R_{earth}$ the radius of Earth, the satellite footprint angle can be calculated as
\begin{equation}
\beta_{max} = \frac{\pi}{2} - \arcsin \left ( \frac{R_{earth}}{R_{earth} + h} \cos \epsilon_{min} \right ) - \epsilon_{min}.
\label{eq:beta}
\end{equation}

A UE at a certain latitude $\lambda$ observes a certain portion of sky above the minimum elevation angle and may connect to a certain number of satellites. The average number of satellites the UE may connect to, denoted by $N_{visible}$, can be calculated as
\begin{equation}
N_{visible} = \frac{N_{sat}}{\pi^2 \sin \alpha} \int_{p}^{q}\sqrt{\frac{\beta_{max}^2 - (x - \lambda)^2}{1 -\left ( \frac{\sin x}{\sin \alpha} \right )^2}} dx
\label{eq:VisibleSat}
\end{equation}
where $p=\max\{-\alpha, \lambda- \beta_{max}\}$ and $q=\min\{\alpha, \lambda + \beta_{max}\}$. The number $N_{visible}$ represents how many satellites a certain UE can connect to at a certain latitude. Equivalently, $N_{visible}$ satellites can be coordinated to cover the region where the UE is located.

We assume that the LEO NTN system deploys Earth-fixed cells, meaning that the cell centers are at fixed locations on Earth and each satellite directs its beams to their cell centers. In order to achieve cells of size similar to the ones formed using UV-plane methodology described in Section \ref{BeamLayout}, we assume that the beams are separated by the distance formed by half of the angle $\phi_{HPBW}$ at nadir direction. With these assumptions, the resulting area of a hexagonal cell, denoted by $A_{cell}$, can be calculated as
\begin{multline}\label{eq:CellArea}
A_{cell} = \frac{\sqrt{3}}{2} R_{earth}^2 \times \\
\left [ \frac{\phi_{HPBW} - \pi}{2} + \arccos \left ( \frac{R_{earth} + h}{R_{earth}} \sin  \frac{\phi_{HPBW}}{2} \right ) \right ].
\end{multline}

The number of cells that are contained in the satellite footprint is calculated by dividing the satellite footprint area by the cell area and is given by
\begin{equation}
N_{cells}^{footprint} = \frac{2 \pi R_{earth}^2 \left ( 1 - \cos \beta_{max} \right )}{A_{cell}}.
\label{eq:CellsInFootprint}
\end{equation}

We further assume that in a certain region of Earth all the satellites that are able to cover the cells in this region divide these cells evenly among themselves. Then the number of cells served by each satellite, denoted by $N_{cells}^{satellite}$, is given by
\begin{equation}
N_{cells}^{satellite} = \frac{N_{cells}^{footprint}}{N_{visible}}.
\label{eq:CellsPerSatellite}
\end{equation}

\begin{figure*}[hbt!]
\centerline{\includegraphics[width=0.7\textwidth]{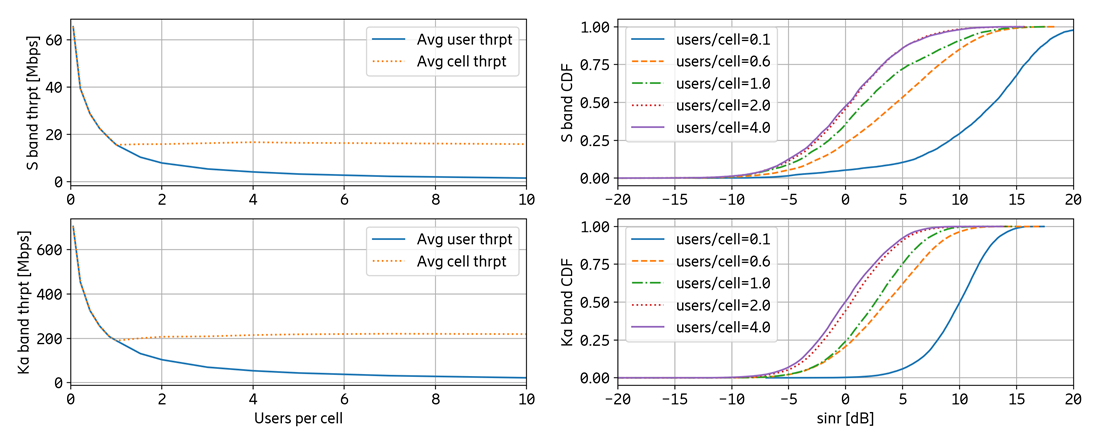}}
\caption{Downlink throughput and SINR for LEO NTN at S band and Ka band.}
\label{f:DL}
\vspace{-4mm}
\end{figure*}

\begin{figure*}[hbt!]
\centerline{\includegraphics[width=0.7\textwidth]{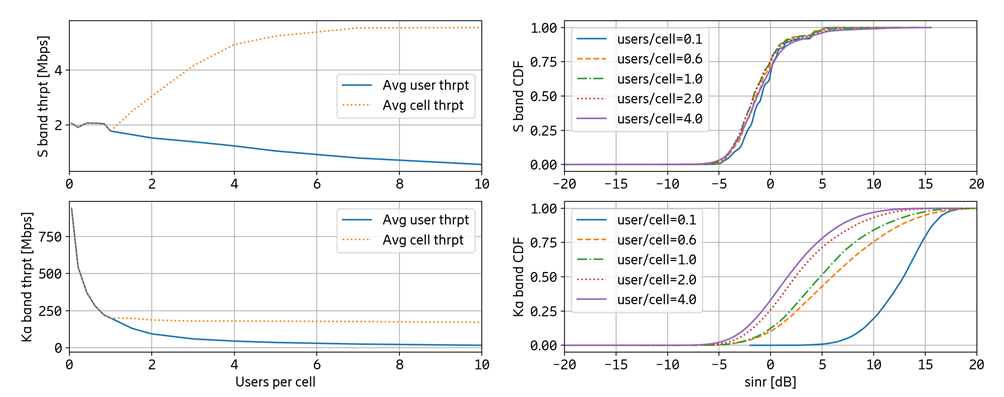}}
\caption{Uplink throughput and SINR for LEO NTN at S band and Ka band.}
\label{f:UL}
\vspace{-4mm}
\end{figure*}

In terms of satellite capacity, we assume that each satellite is able to generate $N_{beams}$ beams in any direction within its footprint at a given time. Each beam is allocated with a certain bandwidth $B$ and two orthogonal polarizations. The capacity offered by a single satellite can be computed as
\begin{equation}
C_{sat} = 2 \eta B  N_{beams}
\label{eq:SatelliteCapacity}
\end{equation}
where $\eta$ denotes the average spectral efficiency. The satellite capacity is shared among all the cells that the satellite is serving. Finally, the capacity per Earth surface area can be computed as
\begin{equation}
C = \frac{C_{sat}}{A_{cell} N_{cells}^{satellite}}.
\label{eq:CapacityDensity}
\end{equation}

\section{System Simulation Results}\label{SimResult}

In this section, we present simulation results on signal-to-interference-and-noise ratio (SINR) and throughput based on the setup described in Section \ref{SimMethodology}. We use full buffer traffic model to get insight about the full system capacity. We study the system performance by varying the user density in the system from 0.1 to 10 users per cell on average. The varying user density creates different system loads and leads to different interference environments.

\subsection{Downlink SINR and Throughput}\label{SimDL}

This section focuses on the downlink SINR and throughput performance, as shown in Figure \ref{f:DL}, where the upper and bottom sub-figures correspond to the LEO NTN operating at S band and Ka band, respectively, whereas the right part shows the downlink SINR, measured at the received physical downlink shared channel (PDSCH) and the left part the user throughput and average cell throughput.

We can see that as the user density increases, the SINR decreases in both S band and Ka band. When the user density is between 0.1 and 1, the average number of UEs per cell is less than the number of cells. In this regime, as the user density increases, inter-beam interference increases rapidly, leading to rapid degradation of the SINR. When the user density increases beyond 2, the system on average approaches fully loaded and enters interference-limited scenario, making SINR insensitive to the user density in this user density regime. 

The cell throughput is the sum throughput of all users in a cell. As the user density increases, the average user throughput decreases because more users share the system capacity and the inter-beam interference increases. Similarly, the average cell throughput decreases and rapidly converges as the user density increases. For the S band, the average downlink cell throughput converges to about 15.6 Mbps. Normalizing the 15.6 Mbps throughput by the 30 MHz system bandwidth in S band yields a spectral efficiency of 0.52 bit/s/Hz. For the Ka band, the average downlink cell throughput converges to about 190 Mbps. Normalizing the 190 Mbps throughput by the 400 MHz system bandwidth in Ka band yields a spectral efficiency of 0.47 bit/s/Hz.

\subsection{Uplink SINR and Throughput}\label{SimUL}

This section focuses on the uplink SINR and throughput performance, as shown in Figure \ref{f:UL}, where the upper and bottom sub-figures correspond to the LEO NTN operating at S band and Ka band, respectively, whereas the right part shows the uplink SINR, measured at the received physical uplink shared channel (PDSCH) and the left part the user throughput and average cell throughput.

We can see that in the S band LEO NTN, the uplink SINR is not high even with very low user density. This is because handheld UE is assumed in the S band LEO NTN and the handheld UE is power limited in the uplink. As a result, the uplink SINR is noise limited and is not much affected by the increased interference resulted from increased user density. In contrast, the Ka band LEO NTN serves VSAT UE, which has high transmit power and is equipped with high gain antenna. Thus, the uplink performance of the Ka band LEO NTN is not power limited. As a result, the corresponding uplink SINR smoothly decreases as the inter-beam interference increases with the user density.

Figure \ref{f:UL} shows that, as the user density increases, the average user throughput decreases because more users share the system capacity and the inter-beam interference increases. Similarly, the average cell throughput of the Ka band LEO NTN decreases and converges rapidly as the user density increases. What is interesting is that the average cell throughput of the S band LEO NTN increases and slowly converges as the user density approaches to 10 users per cell. This is because the handheld UE in S band LEO NTN is power limited and accordingly the scheduler can only allocate a small bandwidth to each handheld UE. As a result, much of the system bandwidth is underutilized when the user density is low. As the user density increases, the scheduler can schedule more UEs in the uplink by increasing the bandwidth utilization in the uplink, leading to increased cell throughput up to the point where the uplink resources are fully utilized.

For the S band, the average uplink cell throughput converges to about 5.4 Mbps. Normalizing the 5.4 Mbps throughput by the 30 MHz system bandwidth in S band yields a spectral efficiency of 0.18 bit/s/Hz. For the Ka band, the average uplink cell throughput converges to about 201 Mbps. Normalizing the 201 Mbps throughput by the 400 MHz system bandwidth in Ka band yields a spectral efficiency of 0.5 bit/s/Hz. 

Table \ref{t:spectralEfficiencies} summarizes the spectral efficiency values presented in this section.

\begin{table}[hbt!]
\vspace{-2mm}
\caption{Average Cell Spectral Efficiency Values}
\begin{center}
\begin{tabular}{lll}
\hline
                  & \textbf{S band} & \textbf{Ka band} \\ \hline
\textbf{Uplink}   & 0.18 bit/s/Hz   & 0.5 bit/s/Hz     \\
\textbf{Downlink} & 0.52 bit/s/Hz   & 0.47 bit/s/Hz    \\ \hline
\end{tabular}
\label{t:spectralEfficiencies}
\end{center}
\vspace{-7mm}
\end{table}

\section{Constellation Capacity Evaluation Results}\label{ConstResult}

In a typical mega-LEO constellation, there may be several thousands of satellites orbiting around Earth. As an example, according to \cite{b15}, the Amazon Kuiper constellation is planned to comprise 3236 satellites. Inspired by this system, we apply the analytical framework presented in Section \ref{ConstMethodology} to analyze a LEO constellation, which comprises 3200 satellites divided into 80 orbital planes with 50° inclination and at the altitude of 600 km. A minimum elevation angle of 35° is assumed for this system. Each satellite can generate 19 beams with different sizes depending on the band considered. Each beam is characterized by two orthogonal polarizations. The system bandwidths are again assumed to be 30 MHz and 400 MHz for S band and Ka band, respectively.

Due to the characteristics of the constellation in question, a user on the ground can be served by up to 25 satellites depending on the user’s latitude. Note that with this constellation, locations above 56° N and below 56° S latitude are out of coverage as no satellites can meet the requirement of 35° minimum elevation angle. 

\begin{figure}[t!]
\centerline{\includegraphics[width=0.8\columnwidth]{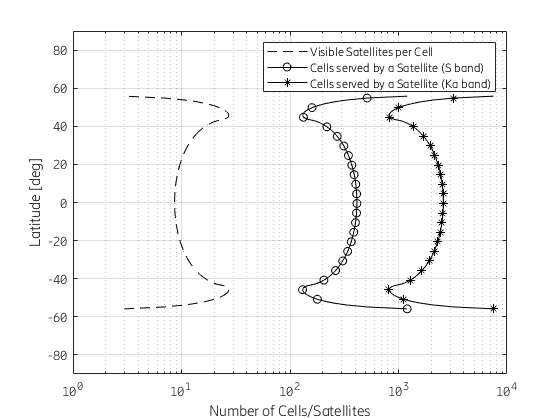}}
\caption{Number of visible satellites and number of cells served per satellite as a function of latitude.}
\label{f:SatNumber}
\end{figure}

The satellite footprint is independent of the band considered. The cell size however depends on the band considered because of different HPBW angles. In particular, it is larger for the S band due to the smaller carrier frequency and larger antenna aperture. This leads to considerably larger cell size and reduces the number of cells in each satellite footprint in the S band. Assuming that the network assigns the cells evenly to all the satellites that can potentially serve them, each satellite will have to handle from 100 to 1000 cells in the S band and from 800 to 8000 cells in the Ka band. These results are illustrated in Figure \ref{f:SatNumber}. The number of cells that each satellite has to serve, in most cases, is considerably higher than the number of concurrent beams that it can generate. This means that each cell is served only for a fraction of time depending on the user positions, the amount of traffic requests, and the scheduling algorithm used by the satellite.

\begin{figure}[t!]
\centerline{\includegraphics[width=0.8\columnwidth]{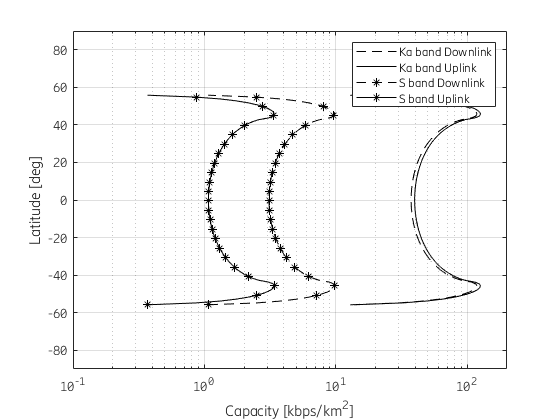}}
\caption{Downlink and uplink capacity density of the LEO NTN.}
\label{f:CapacityDensity}
\vspace{-4mm}
\end{figure}

Figure \ref{f:CapacityDensity} shows the resulting capacity density for all the cases considered. For the S band, the downlink and uplink spectral efficiency values are different (0.52 and 0.18 bit/s/Hz in downlink and uplink, respectively, as summarized in Table \ref{t:spectralEfficiencies}). This leads to a similar asymmetry in the capacity density. The considered constellation is able to support from 0.3 to 3 kbps/km$^2$ in uplink and from 1 to 10 kbps/km$^2$ in downlink depending on the user latitude. 

For the Ka band, the spectral efficiency values are more symmetrical (0.5 and 0.47 bit/s/Hz in downlink and uplink, respectively, as summarized in Table \ref{t:spectralEfficiencies}) so that the capacity density results in similar values for the uplink and downlink in the Ka band, from 14 to 120 kbps/km$^2$. In all cases, each satellite has to serve approximately the same surface area, even if the area may be divided into smaller cells (Ka band) or larger cells (S band). Each satellite using the Ka band, though, is capable of handling a much higher traffic volume due to the larger system bandwidth. Using \eqref{eq:SatelliteCapacity}, the capacity offered by a single satellite in the S band can be computed as 592.8 Mbps in the downlink and 205.2 Mbps in the uplink. Similarly, the capacity offered by a single satellite in the Ka band can be computed as 7.1 Gbps in the downlink and 7.6 Gbps in the uplink.

\section{Conclusions and Future Work}\label{Conclusion}

In this paper, we studied and evaluated the performance of LEO NTN that utilizes 5G NR. The dynamic system simulation was performed based on the latest developments from the 3GPP studies on NTN. We also derived analytical expressions for LEO NTN system capacity estimation. Combining the spectral efficiency values obtained from system simulation and the derived analytical expressions, we showed that the downlink capacity of a LEO satellite in S band with 30 MHz bandwidth serving handheld terminal is about 600 Mbps and the downlink capacity of a LEO satellite in Ka band with 400 MHz bandwidth serving VSAT is about 7 Gbps. We further applied the results to estimate the system capacity of a mega-LEO constellation similar to the Kuiper project proposed by Amazon. We find that due to the large cell sizes in the LEO NTN, the area capacity density is moderate:  1 – 10 kbps/km$^2$ in the S band downlink and 14 – 120 kbps/km$^2$ in the Ka band downlink depending on latitude.

The 7 Gbps capacity offered by a LEO satellite in Ka band with 400 MHz bandwidth may serve fixed wireless access. Assuming an average consumption of 2 Mbps per user, a satellite can serve 3500 users. Since the size of the serving area of a satellite is about $10^5$ km$^2$, the user density is 0.035 users/km$^2$, which may correspond to a rural or remote rural scenario. Another use case of the Ka band satellite may be backhauling for cellular networks. Assuming an average consumption of 50 Mbps per rural cell site, a satellite can serve 140 sites in a rural or remote rural scenario. The 600 Mbps capacity offered by a LEO satellite in S band with 30 MHz bandwidth serving handheld UE is less appealing from user density perspective. With $10^5$ km$^2$ serving area size per satellite and an average consumption of 2 Mbps per user, the user density is only 0.003 users/km$^2$. However, the advantage of the LEO NTN in S band is its capability of connecting handheld devices and providing coverage to areas that are difficult to be covered by terrestrial mobile networks.

This work can be extended in a number of ways. The traffic demand is assumed to be uniform on the ground in the analytical expressions derived in this paper. It is interesting to extend this work to consider more realistic traffic distribution that varies across different areas on the ground. With traffic not uniformly distributed on the ground, it would be of interest to explore efficient scheduling algorithms that a satellite can use to dynamically choose which areas in its footprint to serve. It is also interesting to extend this work to consider techniques that can mitigate inter-beam interference. For example, one could reduce the inter-beam interference by avoiding concurrently scheduling cells that are geographically close.

\end{document}